# Magnetic properties of strained multiferroic $CoCr_2O_4$: a soft X-ray study

Y. W. Windsor[1], C. Piamonteze[1], M. Ramakrishnan[1], A. Scaramucci[2,3], L. Rettig[1], J. A. Huever[4], E. M. Bothschafter[1], A. Alberca[1,5], S. R. V. Avula[1], B. Noheda[4], U. Staub[1]

[1] *Swiss Light Source, Paul Scherrer Institut, 5232 Villigen PSI, Switzerland*

[2] *Laboratory for Scientific Development and Novel Materials, Paul Scherrer Institut, 5235 Villigen PSI, Switzerland*

[3] *Materials Theory, ETH Zurich, Wolfgang-Pauli-Strasse 27. CH-8093 Zurich*

[4] *Zernike Institute for Advanced Materials, University of Groningen, Groningen, The Netherlands*

[5] *University of Fribourg, Department of Physics and Fribourg Centre for Nanomaterials, Chemin du Musee 3, CH-1700 Fribourg, Switzerland*

Using resonant soft X-ray techniques we follow the magnetic behavior of a strained epitaxial film of $CoCr_2O_4$, a type-II multiferroic. The film is [110]-oriented, such that both the ferroelectric and ferromagnetic moments can coexist in plane. X-ray magnetic circular dichroism (XMCD) is used in scattering and in transmission modes to probe the magnetization of Co and Cr separately. The transmission measurements utilized X-ray excited optical luminescence from the substrate. Resonant soft X-ray diffraction (RSXD) was used to study the magnetic order of the low temperature phase. The XMCD signals of Co and Cr appear at the same ordering temperature $T_C \approx 90K$, and are always opposite in sign. The coercive field of the Co and of Cr moments is the same, and is approximately two orders of magnitude higher than in bulk. Through sum rules analysis an enlarged $Co^{2+}$ orbital moment ($m_L$) is found, which can explain this hardening. The RSXD signal of the ($q\ q\ 0$) reflection appears below $T_S$, the same ordering temperature as the conical magnetic structure in bulk, indicating that this phase remains multiferroic under strain. To describe the azimuthal dependence of this reflection, a slight modification is required to the spin model proposed by the conventional Lyons-Kaplan-Dwight-Menyuk theory for magnetic spinels. Lastly, a slight increase in reflected intensity is observed below $T_S = 27K$ when measuring at the Cr edge (but not at the Co edge).

# Introduction

Multiferroics, materials with multiple memory phenomena, have been the subject of intense research over the past decade [1, 2]. This is mostly due to the technological prospects presented by magnetoelectric multiferroics, which possess magnetic and ferroelectric order simultaneously. In this context, significant attention has been given to type-II multiferroics, in which magnetic order drives the electric polarization, such that both order parameters are strongly coupled to each other. Strong enough coupling could allow applications in which magnetization is switched by electric, instead of magnetic fields. This can lead to new energy-efficient applications, because using a voltage to apply electric fields can eliminate the Joule heating associated with the electric current required to generate magnetic fields.

A significant setback in the study of type-II multiferroics is that most known materials of this class possess antiferromagnetic order, and therefore no remnant magnetic moment, a prerequisite for most magnetic applications. Spinel $CoCr_2O_4$ (CCO), which crystallizes in space group $Fd\bar{3}m$ #227, is a rare example of a type-II multiferroic that exhibits ferroelectricity and a remnant magnetic moment in the same phase. CCO is a well-characterized ferrimagnet known since the 1960's [3]. Two well-established magnetic phases are known. A collinear ferrimagnetic phase appears below $T_C \approx 92K$, in which CCO acquires a remnant magnetic moment [3] from non-compensating sublattices of Co and Cr aligned antiparallel to each other along [001]. This is known as the Néel state. Below $T_S \approx 27K$ a spiral component is added to the magnetic structure [3], modulated by an incommensurate $\boldsymbol{Q} = 2\pi(q\ q\ 0)$ modulation vector [4] with $q \approx \tfrac{2}{3}$ [5] ($q$ is slightly temperature-dependent), which results in conical magnetic order. A slight rise in magnetization is also observed in this phase, most likely due to a change in the balance between the magnetizations of the different sublattices. Renewed interest in CCO was sparked upon the discovery of ferroelectricity: the appearance of the spiral is accompanied by the appearance of ferroelectric polarization [6]. Furthermore, strong magnetoelectric coupling is evidenced by the switching of both $\boldsymbol{M}$ and $\boldsymbol{P}$ with a magnetic field [6]. The the appearance of the magnetic spiral and the multiferroic behavior are evidence of strong magnetic frustration. A number of additional magnetic features were reported. A first order transition at $T_L \approx 14.5\ K$ has been reported by several authors [6, 7, 8, 9, 10]. This transition includes a lock-in of the $q$ value, as well a flip in the sign of $\boldsymbol{P}$ [8], and a splitting of the $(q\ q\ 0)$ modulation vector. Several authors have reported an anomaly around 50 K [6, 11, 12, 13]. This feature is often referred to as $T_{kink}$, and may be linked to ordering of Cr moments or a short range ordered spiral.

Theories aimed at describing the magnetic behavior of $AB_2O_4$-type spinels were already introduced several decades ago (in the present case A and B refer to Co and Cr, respectively). Yafet and Kittel [14] proposed a model which considered the exchange coupling between adjacent A or B ions (described by the terms $j_{AA}$, $j_{BB}$ and $j_{AB}$). An important aspect of the model is a stability parameter of the form $u \propto j_{BB}/j_{AB}$, and the system was said to prefer the Néel state when $u \leq 8/9$. An expanded theory known as LKDM [4, 15] (named after Lyons, Kaplan, Dwight and Menyuk) suggested that the conically ordered phase is the ground state when $8/9 < u < 1.298$. The conical order refers to a ferrimagnetic order with a transverse spiral component (i.e. in the plane perpendicular to the ferromagnetic axis). The theory correctly predicted characteristics of the magnetic behavior in these spinels, such as the [110] direction of the modulation vector and the shape of the magnetization's temperature dependence.

Most known multiferroic systems are not suitable for technological applications, usually because the electric polarizations they exhibit are too weak. Therefore the focus in recent years has shifted towards the questions of how to maximize and how to control ferroic orders. To this end, a promising route is the use of epitaxial strain [16]. A number of studies have shown how strain can affect different multiferroic properties. A recent study of CCO films demonstrated that significant variations in magnetic anisotropy can occur due to strain [17]. However the study did not address the multiferroic phase.

In this work we describe a detailed study of the magnetic behavior in strained CCO using resonant X-ray techniques. Resonant X-ray techniques are often employed to study complex oxides. One of their key advantages is element selectivity, which allows separation of signal contributions from one or more constituent ions. CCO is an ideal example of a material where such advantages are required because both the Co and the Cr ions contribute to the macroscopic magnetic behavior. Choi et al. [8] employed resonant X-ray diffraction (RXD) to study single crystals of CCO, focusing mainly on the spiral handedness in the multiferroic phase. Liu et al. [18] employed photoemission and x-ray resonant magnetic scattering around the Co and Cr L edges to characterize epitaxial films.

Here we used RXD to follow the conical order of the multiferroic phase. X-ray magnetic circular dichroism (XMCD) was used to follow the average net moments of the Co and the Cr sublattices in reflection (scattering) and in transmission modes. We specifically employ X-ray excited optical luminescence (XEOL) in transmission to conduct XMCD sum rules analysis. We will show that epitaxial strain can enhance coercivity by 2 orders of magnitude, without cancelling the multiferroic spin spiral (conical) phase.

This paper is divided in the common form, with sections describing the experiments, results, a discussion and a summary. For clarity, the results section is divided into four subsections, each describing a separate experiment.

# Experiments

An 81 nm thick CCO film was grown on a [110]-oriented MgO substrate by pulsed laser deposition (PLD) assisted by reflection high-energy electron diffraction (RHEED). A ceramic $CoCr_2O_4$ pellet made by solid-state reaction was used as a target, from which the material was ablated. Laser pulses from a Lambda Physik COMPex Pro 205-KrF laser with a wavelength of 248 nm, at a frequency of 0.5 Hz were used. The growth took place in a 0.01-mbar oxygen plasma atmosphere, created by an Oxford Scientific mini-electron cyclotron resonance-plasma source in order to improve the film's oxidation. The laser fluence, target-substrate distance and substrate temperature were 3 J/cm^2, 50 mm and 500°C, respectively. Further details regarding post-annealing and substrate preparation are given in Ref. [17]. The precise sample thickness was determined using X-ray reflectivity. Sample crystallinity was characterized by x-ray diffraction at beamline X04SA of the Swiss Light Source (SLS) [19]. The *a*, *b* and *c* lattice constants were determined to be 8.292 Å, 8.294 Å, and 8.383 Å, corresponding to strain values of 0.48%, 0.45%, and -0.61%, respectively (positive values denote compressive strain). From these values we find that the [110] direction is 0.9% strained, and almost no strain is present along the [1$\bar{1}$0] direction (~0.02%). A three-dimensional depiction of the conventional unit cell is presented in Figure 1a. A small deviation from 90° was observed in the angle between the [100] and [010] directions ($\gamma = 90.5°$). The effect of this deviation is presented in Figure 1b and Figure 1c, which depict side views of unstrained and strained unit cells.

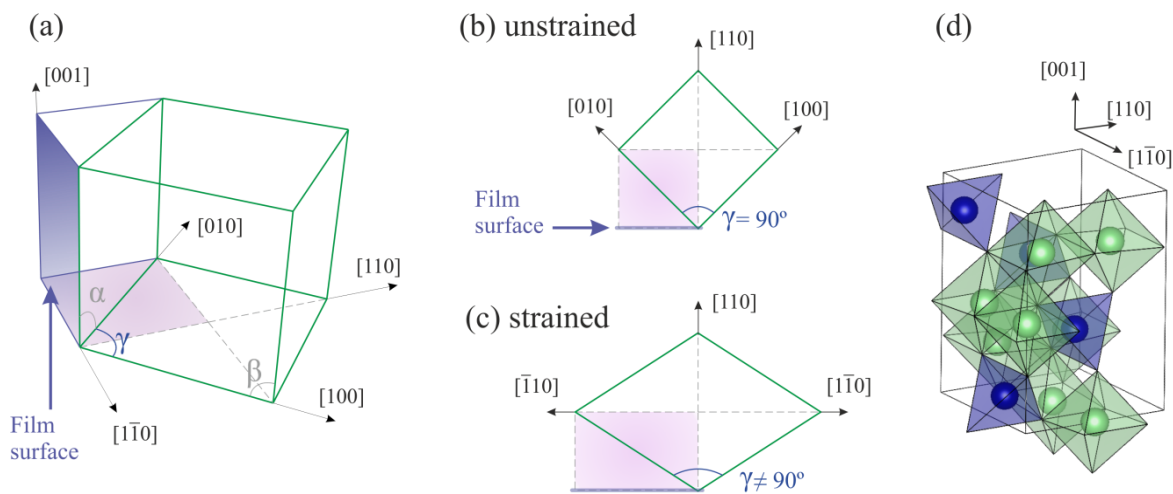

Figure 1 – geometry of strained [110]-oriented $CoCr_2O_4$. (a) A 3D view of the unit cell in cubic setting. The dark shaded (blue) area represents the film's surface. The lightly shaded (pink) area represents the square enclosed by diagonals. (b) and (c) represent the 2D projection of the cubic unit cell onto the *ab* plane, in the unstrained and the strained cases, respectively. The lightly shaded (pink) area from (a) is also shown here to emphasize that the diagonals remain orthonormal. The film surface is also shown as a thick (blue) line. (d) Alternatively defined unit cell of $CoCr_2O_4$ in a tetragonal setting, containing four formula units. This unit cell becomes orthorhombic when strained. The pink area in (b) and (c) are actually projections of this unit cell.

Bulk CCO and MgO are both cubic. Therefore growing CCO on [110]-oriented MgO can lower the symmetry of CCO due to lattice mismatch, which could produce a number of different structural domains. In bulk CCO, the macroscopic observables *M*, *P* and *Q* are all perpendicular to each other, being along [1$\bar{1}$0], [001] and [110] or other equivalent axes, respectively [6]. We have not observed any significant ambiguity in the directions of the macroscopic observables[1], and therefore we do not consider other structural domains besides the [110] surface orientation. This orientation sets both macroscopic observable quantities (*P* and *M*) in plane and perpendicular to each other, and it sets *Q* out of plane. Bulk CCO can be described by a tetragonal unit cell which contains 4 formula units, instead of 8 in the cubic setting. This unit cell is defined by the same [110], [1$\bar{1}$0] and [001] directions (see figure 1d). Figure 1c illustrates that these directions remain orthonormal even when $\gamma \neq 90°$. The tetragonal cell thus becomes orthorhombic[2]. To avoid confusion, all crystallographic directions in this paper are given in the cubic notation.

Magnetization measurements were conducted using a commercial MPMS SQUID magnetometer equipped with a 7 T magnet. Substrate contributions were not subtracted from the data. In field dependent magnetization measurements, linear contributions to magnetization above saturation were subtracted. X-ray scattering measurements at the Co and Cr $L_{2,3}$ edges (~780 eV and ~590eV, respectively) were conducted using the RESOXS [20] high vacuum diffractometer at beamline X11MA of the SLS [21]. Data were collected using an IRD AXUV100 photodiode covered by a 400 nm thick aluminum foil for blocking visible light. The scattering geometry is schematically shown in Figure 2a. X-ray excited optical luminescence (XEOL) measurements were conducted at the Co and Cr $L_{2,3}$ edges using the high-field XTreme end station at beamline X07MA of the SLS [22]. Transmission spectra were recorded using XEOL by correcting for the energy dependence of luminescence from the bare substrate, as demonstrated by Kallmayer et al [23]. Other authors have since employed XEOL [24, 25, 26, 27], and the luminescence of bulk MgO was studied in detail by Vaz et al. [28]. The spectra were recorded by continuously scanning the incoming photon energies at a rate of 0.44 eV/s and 0.36 eV/s around the Co and Cr resonances, respectively. The cold head's sample mount was designed to accommodate a photodiode *behind* the sample. The experiment was conducted with the substrate mounted in front of a small hole, through which light from the substrate can reach a photodiode. The hole is the only opening of the photodiode region. For sum rules analysis data were binned using 0.25 eV steps, which is well above the energy resolution of this experiment. Reducing the bin size even by a factor of 2 does not appreciably affect the results.

---

[1] As shown in later sections, the [001] is the easy magnetization axis, and the ordering wave vector **Q** is along [110].

[2] For this we also assume that the difference between the *a* and *b* lattice constants is negligible, as it is close to the limit of our experimental resolution,

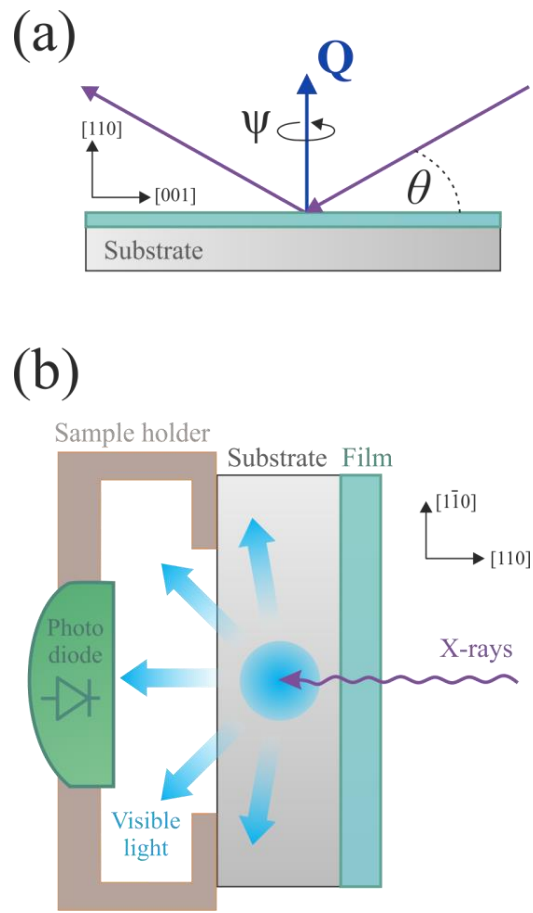

Figure 2 – Schematics of the soft X-ray experimental setups. (a) The scattering setup. Ψ is the rotation angle around the momentum transfer vector Q, and Ψ = 0° is defined when the [001] direction is in the scattering plane (directions shown correspond to this orientation). (b) The XEOL setup. The sample is mounted on a cold copper piece with a ~3mm hole through which the substrate can luminesce to the photodiode.

# Results

1. Magnetization

Figure 3a presents the temperature dependence of magnetization measured upon warming along the [001] direction and along [110], the film normal. Before the measurements the sample was cooled under a field of 2 T, and a field of 10 mT was applied during the measurements. We find that the magnetization along [001] is nearly 50% higher than in bulk (compared to Ref. [6]). The sharp rise at the lowest temperatures is most likely due to a Curie-Weiss behavior of substrate impurities. This contribution is strongly suppressed upon warming.

Figure 3b presents two magnetization curves measured along [001] as function of magnetic field at different temperatures. Similar measurements along [110] exhibited no hysteresis (not shown). Direct comparison with bulk data such as in Ref. [8] reveals that the coercive field is two orders of magnitude larger in our strained film. A similar observation was made for a [001]-oriented strained film in Ref. [17]. At 5K and 15K the 7 T field available was unable to saturate the sample. Data were also taken out of plane, along [110] (not shown), and no clear hysteresis or saturation were observed, as expected for a hard axis.

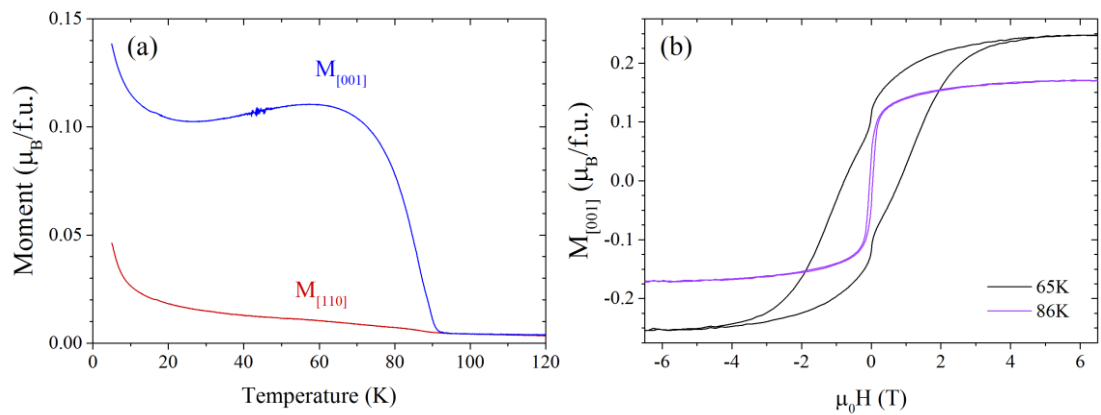

Figure 3 – Magnetization of the CCO sample. (a) Magnetization as function of temperature, measured upon warming along [001] and along [110]. Measurements were done under a field of 10 mT after field cooling in 2 T (fields were applied along the measurement direction). The noise at ~45 K is most likely an experimental artefact. (b) Magnetization as function of magnetic field along [001], measured at 65 K and 86 K. Linear contributions were removed above 6T, which is above the saturation field of the hysteresis (see main text).

The data in Figure 3 suggest that the [001] axis is the easy magnetization axis, as in bulk. To obtain further insights on the favored direction of the spins we performed *ab initio* calculations using LSDA+U as implemented in the Vienna Ab Initio Simulation Package [29], in a similar manner as done in Ref. [17]. We used a projector augmented wave pseudopotential [30] and fixed the strength of the effective on-site Coulomb interaction and effective Hund's rule coupling to $U_{Co}$=4 eV, $J_{Co}$=1 eV and $U_{Cr}$=3 eV, $J_{Cr}$=1 eV for $Co^{2+}$ and $Cr^{3}$, respectively. For simplicity, in the following calculation, energies are derived for the tetragonal unit cell of bulk $CoCr_2O_4$ shown in Figure 1d. For such settings we use a 9 x 9 x 5 Monkhorst k-points mesh. Note that energies calculated in Ref. [17] were for a unit cell with only two formula units.

The relaxed (i.e. bulk) unit cell acquired through LSDA+U has a lattice constant of 8.212 Å in the cubic setting. To determine the crystallographic structure in the film we fixed the in-plane lattice constants: the [1$\bar{1}$0] was fixed to its size in the relaxed cell, and the [001] was set to 1.006 times its size in the relaxed cell (0.6% tensile strain). We then relaxed the out-of-plane lattice direction [110] and the ionic positions within the unit cell. The [110] out-of-plane length of the resulting strained unit cell was determined to be 11.582 Å (in cubic setting). The relaxation was performed in absence of spin-orbit couplings and keeping the spins of $Cr^{3+}$ antiparallel to those of $Co^{2+}$. Interestingly, the *ab initio* relaxation results in a reduction of approximately 0.3% in the out-of-plane length along the [110] direction (with respect to the bulk value). While it maintains the correct direction and order of magnitude, this underestimates the experimentally observed 0.9% reduction.

To estimate the easy axis of magnetization, the energy of different spin configurations was calculated. In each configuration the spins' direction was fixed at an angle $\phi$ between two crystallographic axes in the strained unit cell (described above). Figure 4 presents results for two rotations: between the [001] and [110] directions (black circles), and between the [001] and [1$\bar{1}$0] directions (blue circles).

From these calculations we conclude that the minimal energy occurs for spins oriented along [001]. Furthermore, the energy difference between [001] and the other two axes is nearly the same. In other words, the [1$\bar{1}$0] axis is the hard axis in-plane, and is nearly as hard as the out-of-plane [110] direction.

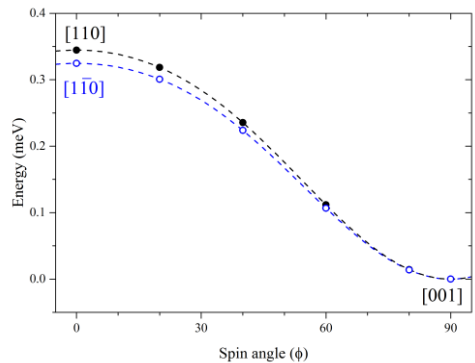

Figure 4 – Energy dependence of different spin arrangements. In each arrangement the spins are set along an axis at an angle $\phi$ between the two crystallographic directions. Solid circles (black) indicate spins along an axis between [001] and [110]. Open (blue) circles indicate spins along an axis between [001] and [1$\bar{1}$0]. Dashed lines are the best fits (see main text). In all cases the $Cr^{3+}$ spins are kept antiparallel to the $Co^{2+}$ spins. Calculations were done for a unit cell with four formula units, and the energy of $\phi = 90°$ (spins along [001]) is subtracted for clarity.

The calculated energy dependence fits well to the dependence of anisotropy described in Ref. [17] as $E(\phi) = \alpha \cos^2 \phi + \beta [\cos^4 \phi + \sin^4 \phi]$ (see dashed curves in Figure 4). The parameters $\alpha$ and $\beta$ describe second and fourth order terms of the spin anisotropy from all magnetic sublattices. The best fit parameters are listed in Table 1.

Table 1 – Best fit parameters to the dependence of energy on spin rotation, following Ref. [17], for the energy values calculated and shown in Figure 4 for a unit cell with 4 formula units. Bulk values are from [17].

| Rotation direction | $\alpha$ (μeV) | $\beta$ (μeV) |
|---|---|---|
| $[110] \rightarrow [001]$ | 344.56 | −68.80 |
| $[1\bar{1}0] \rightarrow [001]$ | 324.61 | −68.45 |
| Bulk CCO (cubic) | 0 | −44.5 |

2. Resonant X-ray scattering

Resonant X-ray scattering is an element-selective probe of a material's electronic state [31], including its magnetic configuration. For transition metals such as Co and Cr, electric dipole (E1) excitations at the $L_2$ and $L_3$ absorption edges are direct probes of the unoccupied 3d electronic levels. We begin by employing X-ray magnetic circular dichroism (XMCD) in reflectivity mode, to follow the net moments of the Co and Cr ions separately. Figure 5a and b present the energy dependence of scattered intensity around the Co and the Cr $L_{2,3}$ edges, measured using either left- or right-handed circularly polarized incoming light. We refer to the intensity measured from scattering circularly polarized incident light as $I_{\pm}$, where $\pm$ indicates the handedness of the incoming light. The experimental configuration is such that the [001] crystal axis, the easy axis of magnetization, is in the scattering plane, as in Figure 2a. The measurement was done at 10K, and the sample was cooled prior to the measurement in a field of 450 mT or -450 mT applied along [001]. A clear contrast is observed between the spectra measured with $I_+$ and $I_-$. The incident angles of the light on the sample were shallow ($\theta = 4.1°$ and 7.3° for Cr and Co, respectively), and were chosen because they exhibited largest circular dichroism.

Panels (c) and (d) of Figure 5 present a temperature dependence of $I_+$ and $I_-$ at the Co and Cr $L_3$ edges. The energy with the highest contrast between $I_+$ and $I_-$ were used: 779.15 eV and 577.9 eV, marked by vertical lines in panels (a) and (b). The circular dichroism disappears above $T_N \approx 90K$, indicating that the dichroism is indeed due to magnetic order (XMCD). Furthermore, the reflectivity at the Cr edge rises by ~5% below ~27K (for both circular polarizations), which can be attributed to the appearance of conical magnetic order, known to occur in bulk CCO below $T_S = 27K$ (indicated by a dashed line in Figure 5c).

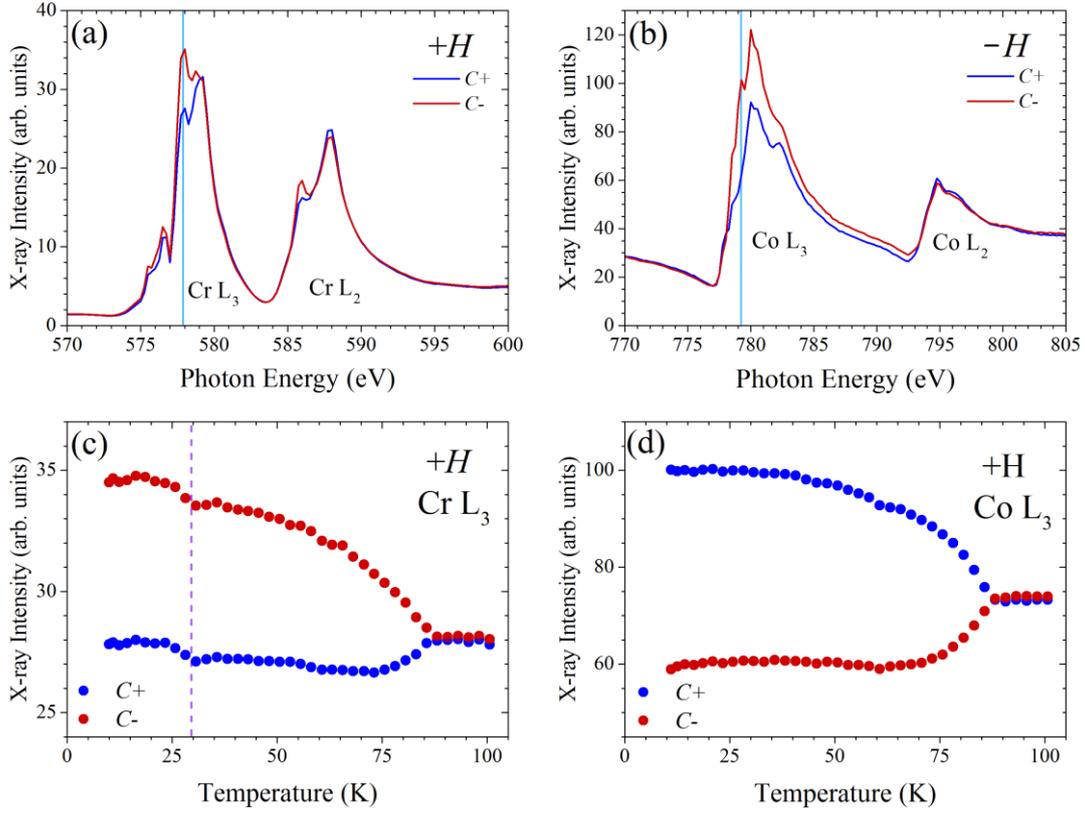

Figure 5 - Reflectivity data taken from CCO using circular light after field cooling. (a) and (b) are reflectivity signals as functions of incoming photon energy around the Co and Cr L edges, respectively. Data were taken at 10K, and at incident angles of $\theta = 4.1°$ and $7.3°$ for Cr and Co, respectively. Vertical lines represent the energies used in the panels (c) and (d). Panels (c) and (d) present reflectivity at the $L_3$ edges as functions of temperature. Prior field cooling was conducted in $\pm H = \pm 450 \, mT$. Note that panel (b) presents data taken after field cooling in –H, while the other panels were taken after cooling in +H. Error bars are smaller than the symbols and lines.

For further analysis of the measured intensity, the resonant magnetic scattering amplitude in the case of an E1-E1 event can be written as [32]

$$f_{E1E1} \propto \{(\hat{\varepsilon}'^* \cdot \hat{\varepsilon})F^{(0)} - i\hat{m} \cdot (\hat{\varepsilon}'^* \times \hat{\varepsilon})F^{(1)} + (\hat{\varepsilon} \cdot \hat{m})(\hat{\varepsilon}'^* \cdot \hat{m})F^{(2)}\} \quad (1)$$

Here the magnetic moment is represented by $\hat{m}$, the terms $F^{(n)}$ represent prefactors, and $\hat{\varepsilon}$ and $\hat{\varepsilon}'$ are the incoming and outgoing polarization vectors. For magnetic scattering, the second term is of primary importance, as it is linear with respect to the magnetic moment. In order to maximize sensitivity to $\hat{m}$, the direction of incoming light should be almost parallel to the magnetic axis. This is equivalent to the notion that XMCD in absorption or transmission experiments is sensitive to the moment parallel to the incoming beam. In the present case, the magnetic easy axis is in the film plane (along [001]), so high sensitivity can be achieved with a small incident angle $\theta$, and when the film is oriented such that [001] is in the scattering plane (as in Figure 2a).

Scattered intensity will generally follow $\propto |f_0 + f_{E1E1}|^2$, so any signal that is linearly proportional to $\hat{m}$ can only arise from interference between the first and second terms of Eq. (1) ($f_0$ is the non-resonant scattering amplitude, and it interferes in the same manner). From a detailed inspection of Eq.

(1) with the experimental geometry described above, we find all such linear contributions go as $\propto \cos\theta$ or as $\propto \cos\theta \cos 2\theta$. The non-magnetic contributions go as $\propto \cos 2\theta$ and $\propto \cos^2 2\theta$, and the $|F^{(1)}|^2$ second-order contribution (i.e. proportional to $\hat{m}^2$) goes as $\propto \cos^2\theta$. All other contributions to intensity from Eq. (1) include powers of $\sin^2\theta$, and are therefore negligible in the limit of small $\theta$. Therefore, the intensity linearly proportional to $\hat{m}$ is the only significant term expected to produce circular dichroism in the limit of small $\theta$.

The data in panels (c) and (d) of Figure 5 indicate that the resonant non-magnetic term (first term in Eq. (1)) is the largest contribution to the resonant signal. The XMCD, which comes from the interference between the first and second terms of Eq. (1), is the second major contribution, representing up to ~20% of the measured signal. In the Cr data presented in Panel (c), the 5% rise of intensity below 27K bears the same sign in $I_+$ and $I_-$. It is therefore either due to the second order magnetic term (second term in Eq. (1)), or due to a change in the non-magnetic term of Eq. (1). This feature is notably absent in the signals from the Co edge in Panel (d).

The data in Panels (b) and (d) of Figure 5 were both taken around the Co L edges, but after field cooling in opposite fields. The circular dichroism contrast between $I_+$ and $I_-$ is therefore reversed in the two panels. The same reversal effect occurs at the Cr edge. To easily follow this effect, we define magnetic *asymmetry* as $(I_+ - I_-)/(I_+ + I_-)$. Figure 6 presents asymmetry calculated from the temperature dependence. Panels (a) and (b) correspond to data taken after field cooling in 450 mT and -450 mT, respectively. The results indicate a complete reversal of the Co and Cr asymmetry with opposite field cooling, and that the moment directions of Co and Cr are clearly opposite to each other. This is in agreement with the models presented in Ref. [13]. Also, worth noting is the absence of any significant change around $T_S$, indicating that the rise observed in Figure 5c below $T_S$ does not contribute to circular dichroism, and can therefore arise from non-magnetic terms, or from the term proportional to an even power of $\hat{m}$ (e.g. $\hat{m}^2$), as mentioned above. The latter could imply that a change in X-ray magnetic linear dichroism (XMLD) occurs at $T_S$, which is a strong effect for certain multiplet features in antiferromagnetic oxides [33, 34]. However, the total Cr moment is not known to change in size below $T_S$, so a change in XMLD is likely to be accompanied by a change in XMCD, which is not observed.

The magnitude of asymmetry is different for Co and Cr, but this depends largely on experimental parameters such as the incident photon energies used. However, the temperature dependence of asymmetry from the two sublattices is also qualitatively different. Figure 7 presents all four asymmetry curves from Figure 6, normalized to their value at 21 K (this temperature was chosen because both curves are nearly flat there). These normalized curves for Co and Cr differ most around ~70 K. This is in agreement with the magnetization measurements (see Figure 3a), which show that the total moment is highest in this temperature region [6], assuming that the Co and Cr moments are similar at low temperatures. A second observation is that although all four field coolings were conducted under the same field strength, the gap between Co and Cr is different in magnitude for the two opposite field directions (open and solid symbols). This may indicate the existence of an exchange bias, or it could be an experimental artifact due to the manual use of a permanent magnet for applying the field.

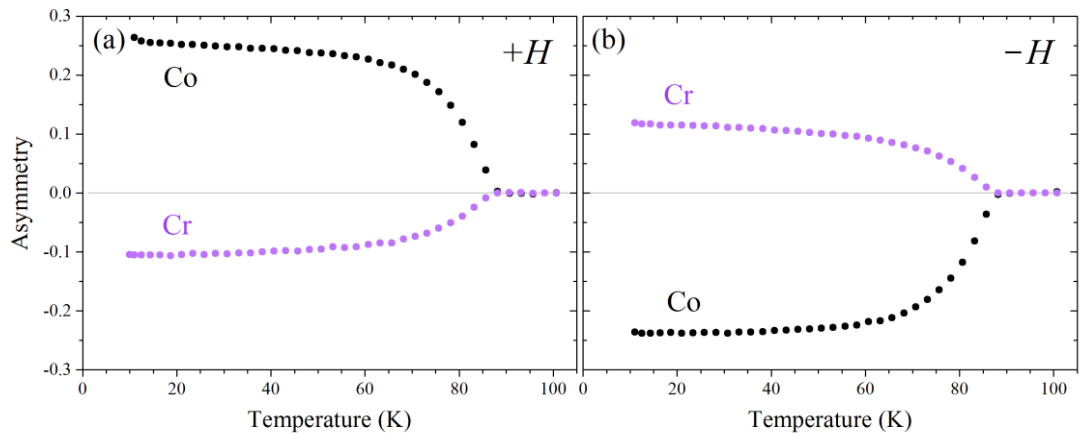

Figure 6 – Magnetic asymmetry as a function of temperature measured at the Co and Cr L$_3$ edges (779.15 eV and 577.9 eV). The left and right panels present four heating cycles conducted with the exact same procedure, but after cooling in magnetic fields of opposite signs: -400 mT and 400 mT, applied along [001]. Error bars are smaller than the symbols.

It is also worth noting that below ~16 K, the data indicate that a split may occur, as shown in the inset: for field cooling in $-H$, the Cr magnetization grows further upon cooling, while for $+H$ the Co magnetization grows further upon cooling (corresponding to the same moment direction). This coincides with the temperature at which an additional first-order magnetic transition is known to occur in bulk CCO [7, 8]. This effect cannot be explained by a difference in the magnitude of applied field, as the effect is the same for opposite fields. It can be explained as the result of an exchange bias, possibly coming from the interface to the substrate.

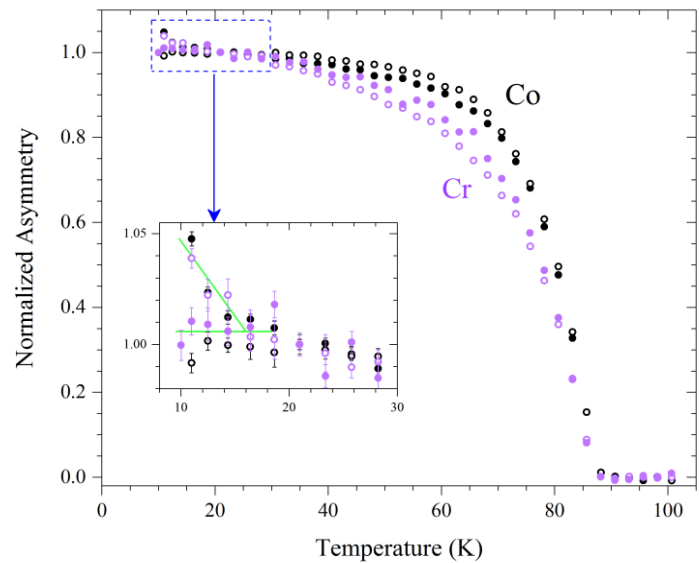

Figure 7 – Normalized asymmetry as a function of temperature. All curves from Figure 6, normalized to their value at 21 K for direct comparison. Solid (open) symbols are data taken after cooling in 450 mT (-450 mT). The inset is a close-up of the low temperature region (lines are guides for the eyes).

3. Resonant soft X-ray diffraction

Resonant soft X-ray diffraction has been frequently employed in recent years to study antiferromagnetic structures of thin films [35, 36], and particularly multiferroic films [37, 38, 39]. This is because the large resonant enhancement of magnetic diffraction even from very small sample volumes [40].

A signature of the low-temperature multiferroic phase in CCO is the appearance of a spiral component in the magnetic structure below $T_S = 27K$. Combined with the high-temperature ferrimagnetic arrangement, this results in a conical magnetic structure [3]. The spiral is described by a magnetic wave vector of $\boldsymbol{Q} = (q\ q\ 0)$, with $q \approx 2/3$. We now aim to observe the spiral using RSXD by fulfilling Bragg's law at the Co $L_3$ edge. This cannot be done at the Cr L edges, because they are too low in energy. Since the [110] direction is perpendicular to the sample surface, the same experimental conditions of the previous section can be kept, except the incident angle is now changed to the Bragg angle $\theta = \sin^{-1}(hc|\boldsymbol{Q}|/2E)$, in which $E$ is the incoming photon energy, $h$ is Planck's constant and $c$ is the speed of light, and $\boldsymbol{Q}$ is in units of rad Å$^{-1}$. Scattering is now expected to arise solely from the 2$^{nd}$ term in Eq. (1). Summing up all individual contributions from moments in the magnetic super cell, one can write the magnetic structure factor as

$$F_{E1E1}(\Psi) \propto (\hat{\varepsilon}'^* \times \hat{\varepsilon}) \cdot \sum_i \widehat{m}_i(\Psi) e^{2\pi i \boldsymbol{Q} \cdot r_i} \quad (2)$$

in which $r_i$ is the position of the ion carrying moment $\widehat{m}_i$. $\Psi$ is the azimuthal angle (see Figure 2a), which is defined as 0° when [001] is in the scattering plane.

Figure 8 presents the temperature dependence of scattered intensity from the ($q\ q\ 0$) reflection at the Co $L_3$ edge, measured upon warming with the sample oriented such that the [001] direction is in the scattering plane ($\Psi = 0°$). The appearance of this reflection indicates that the film exhibits the low temperature (multiferroic) phase despite the high strain. Values of $q$ in reciprocal lattice units (r.l.u.) are calculated using the lattice constants measured at room temperature and the size of the momentum transfer, shown in the top horizontal axis. The right-hand edge of the data is limited by the edge of accessible reciprocal space at this energy. The magnetic intensity continuously changes within the accessible temperature range, and disappears near the expected value of $T_S$. The energy dependence of this reflection is presented in Panel (b). The simple energy profile that emerges suggests that most of the intensity is from a few dominant multiplet transitions around the Co $L_3$ edge.

Panel (c) presents the integrated intensity of the reflection, which continuously decays upon warming. Linear extrapolation of the curve suggests $T_S \approx 27K$, as seen in bulk [3]. Short range correlations are found also above this temperature. The modulation parameter $q$, presented in Panel (d), changes in the rage 0.63 – 0.69 r.l.u., which includes all previously reported $q$ values: 0.63 [3], 0.67 [7] and 2/3 [8]. Unlike other reports on bulk samples, here $q$ does not lock in to a constant value [7, 8], so no indication of the first order transition around 14.5 K is observed. The correlation length of the spiral, presented in Panel (e) and calculated as $\zeta = 2/FWHM$ (FWHM indicates the full width at half maximum), does not change considerably throughout the measured temperature range, and remains at around 10-12 nm, suggesting that the probed order is short-ranged, but approximately 3 times longer than in a single crystal [5]. The large range of $q$ in panel (d) and the broad width of the reflection may indicate that the observed reflection overshadows weaker reflections, such as those which would indicate the existence of the 15K transition. Lastly, no $(\delta\,\delta\,0)$-type reflections with $\delta = 1 - q$ have been observed at the Co or Cr edges, although these were predicted in Ref. [41].

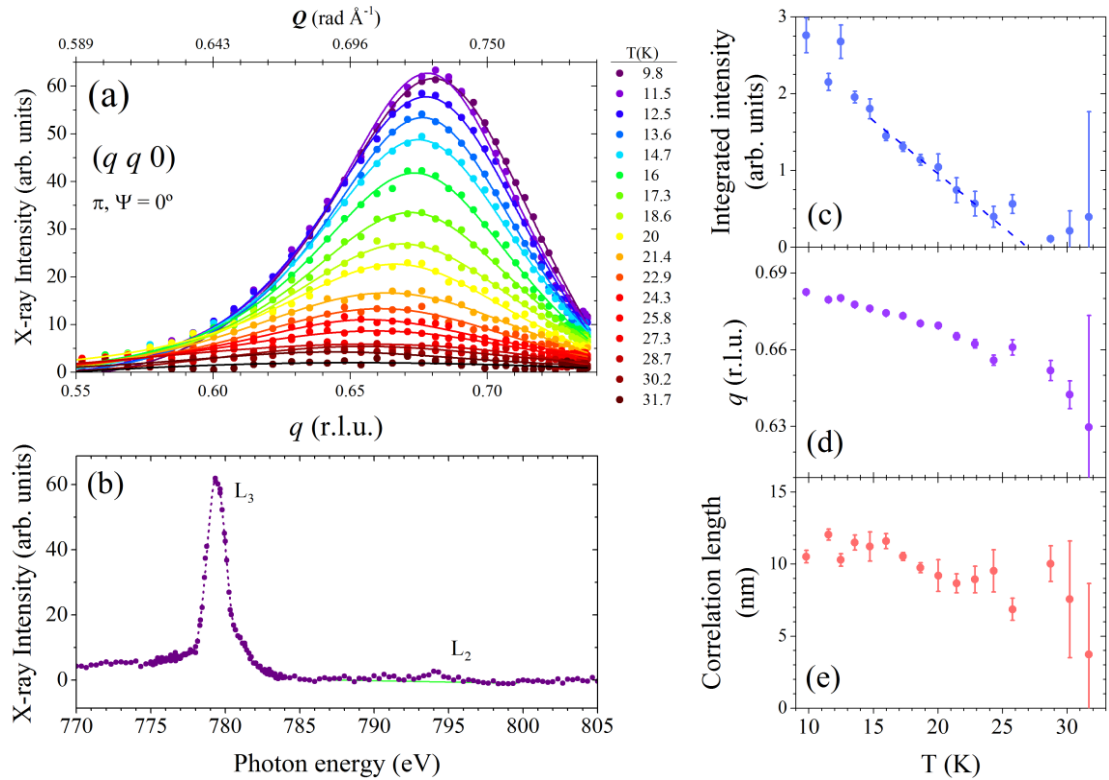

Figure 8 – Temperature of the $(q\,q\,0)$ reflection from a [110]-oriented $CoCr_2O_4$ film. (a) Data taken upon warming, using incoming $\pi$ polarized linear light with an energy of 779.2 eV (Co $L_3$) with the sample at an azimuth of $\Psi = 0°$ ([001] axis in the scattering plane). Lines are best fits of the data to Voigt profiles. Scans were conducted along the [110] (specular) direction, so the horizontal axis is the corresponding $q$ value in $(q\,q\,0)$. The upper horizontal axis presents the actual magnitude of the momentum transfer $\mathbf{Q}$. (b) Energy dependence of the integrated intensity of the $(q\,q\,0)$ reflection, taken using incoming $\pi$ polarized linear light with the sample at 10K and at an azimuth of $\Psi = 0°$. (c) (d) and (e) present the temperature dependence of the integrated intensity, the modulation parameter $q$, and the correlation length, respectively. The Lines in (b) and (c) are guides for the eye.

To further study the spin structure, an azimuthal dependence of the $(q\,q\,0)$ reflection was conducted at 10K, by rotating the sample by an angle $\Psi$ around the [110] axis. The azimuthal dependence of the integrated intensity is presented in Figure 9 for incoming $\pi$ and $\sigma$ linearly polarized light.

To describe the observed azimuthal dependence, we follow LKDM theory [4] as it successfully reproduces many observations. The theory describes the system as six magnetic fcc sublattices, numbered by $\nu = 1\,...\,6$. (two $A$ fcc sublattices at the 8a diamond sites, four $B$ fcc sublattices at the 16d pyrochlore sites). The $i$th moment in the sublattice is described [8] as:

$$\boldsymbol{S}_{i\nu} = \sin\varphi \left[\hat{x}\cos(\boldsymbol{\tau}\cdot\boldsymbol{r}_{i\nu} + \gamma_\nu) + \hat{y}\sin(\boldsymbol{\tau}\cdot\boldsymbol{r}_{i\nu} + \gamma_\nu)\right] + \hat{z}\cos\varphi \qquad (3)$$

This represents a spiral in the *ab* plane, plus a constant moment along *c*. Here $\boldsymbol{\tau}$ is the ordering wave vector, $\boldsymbol{r}_{i\nu}$ is the position of the *i*th atom in sublattice number $\nu$, $\gamma_\nu$ is the phase of the spiral on sublattice $\nu$, and $\varphi$ is half the opening angle of the cone (i.e. the canting angle away from the $\hat{z}$ axis that the spiral causes). Since our data were taken at the Co $L_3$ edge, we calculate the expected azimuthal dependence from this magnetic motif by plugging only the two Co sublattices into Eq. (2). The results are presented as dashed lines in Figure 9. Clearly this model does not agree with the measured data.

A possible reason for this discrepancy was recently suggested by Macke et al. [42], who demonstrated that dynamical diffraction effects can account for discrepancies between measured data and expected azimuthal dependences calculated in the kinematic limit. However, given that the present system is a thin film with a limited probed volume measured in specular geometry (implying that the azimuthal rotation does not affect the beam's path length in the sample), we believe this is not the case here.

To reconcile the LKDM model and the data, we find that only one parameter needs to be modified. LKDM theory [4] defines that the two Co sublattices have the same $\gamma_\nu$ phase in their spiral components. The solid line in Figure 9 represents a calculation for the same model, but with a phase of $\pi/4$ between the two Co sublattices. A reasonable agreement with experimental data is reached, and we conclude that the spiral state is not entirely the same as in the bulk picture presented by LKDM theory. A possible explanation for the change can be the temperature (10 K), which is below the anomalous ~15 K magnetic transition. This transition is known to affect the magnetic structure, and is not described by LKDM theory.

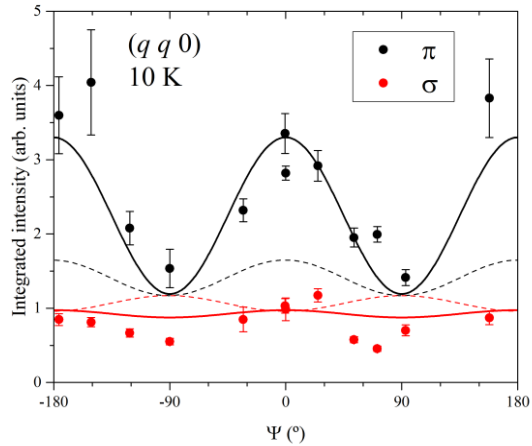

Figure 9 – Azimuthal dependence of the (*q q* 0) reflection in CCO, measured at 10 K using incoming light with either $\sigma$ or $\pi$ linear polarization. Solid lines are calculations with a phase of $\pi/4$ between the Co sublattices. The dashed lines are a calculation with 0° phase, as in LKDM theory. Calculations assumed $q = \tfrac{2}{3}$ and used Eq. (2) to calculate the structure factor.

4. XMCD study using X-ray excited optical luminescence

The main setback in conducting XMCD experiments in reflectivity mode, as previously described, is that it is difficult to obtain quantitative information. One of the major advantages of XMCD is sum rules analysis, which can provide a quantitative estimate of spin [43] and orbital [44] moments. This is commonly used in XMCD experiments conducted in absorption or transmission modes.

We now aim to employ sum rules analysis to quantify the average moments on the Co and Cr sublattices. However, as the film is an insulator, conventional electron yield is not possible. Conventional transmission experiments are also not possible, since the substrate is 0.5 mm thick. As an alternative to these methods, we use the sample's substrate as a scintillator. For this we utilize the intense X-ray excited optical luminescence (XEOL) of the substrate, which is a measure of the intensity of the beam transmitted through the thin film. XEOL is not a commonly used method, but it has been successfully employed before [45], both as an XMCD probe [26] and specifically for sum rules analysis [23]. The efficiency of XEOL varies with energy. To account for this, the transmitted intensity as a function of the sample's thickness can be described as:

$$I(z) = I_0 \Lambda(E) e^{-\mu z} \tag{4}$$

Here the term $\Lambda(E)$ is the efficiency function of the XEOL for the substrate used, which must be measured and accounted for prior to any further analysis. The other terms, $I_0$ and $\mu$, represent the incident intensity and the absorption coefficient, respectively. The substrate is thick enough to absorb the entire transmitted X-ray beam (the X-ray attenuation length of MgO at these energies is between 0.3 and 0.7 $\mu m$). We therefore assume that the measured signal is entirely XEOL. Figure 10 presents raw uncorrected XEOL spectra around the Co L edges, taken from our CCO sample using circularly polarized incoming light of both helicities. Clear circular dichroism is observed.

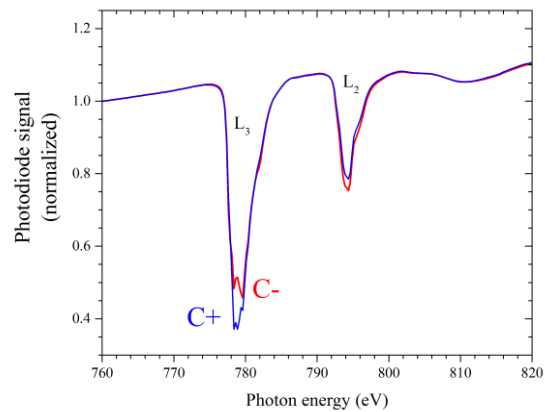

Figure 10 - XEOL data from CCO around the Co L edges. CCO data are taken with the sample at 15.5K, and using opposite helicities of circularly polarized incoming light. The spectra are normalized to their value at 760 eV.

We now detail the application of sum rules analysis to the XEOL data, mainly following Chen et al. [46]. The sample was oriented such that the X-ray beam was 30° from the [001] axis. Before measurements, the sample was cooled from 150 K to 2 K in a 2 T field along the beam direction.

Figure 11 presents data taken at 15.5K. Panels (a) and (b) present the XAS ($\mu_+ + \mu_-$) and XMCD ($\mu_+ - \mu_-$) spectra taken around the Co $L_{2,3}$ edges. The absorption coefficients from either helicity ($\mu_\pm$) were corrected for $\Lambda(E)$ before addition/subtraction. The right hand axes represent integrals over these spectra. Panels (c) and (d) present the same spectra, taken around the Cr $L_{2,3}$ edges.

For determining moment sizes, the integrals are evaluated at an energy above the $L_2$ edge in which their profile flattens out (energies chosen for Co and Cr were 801 eV and 598 eV, respectively, marked as dashed lines in all panels). Beyond these energies pronounced EXAFS oscillations appear in the XAS spectra in panels (a) and (c). These can be reproduced using *ab initio* simulation code such as FDMNES [47] without considering any magnetic order (not shown). The integral over the XAS spectrum was taken after a two-step function is subtracted, which represents the continuum contribution to the $2p \rightarrow nd$ E1 transitions at the $L_{2,3}$ edges. The nominal 2:1 branching ratio was used to define the relative height of each step. This is the ratio between the probability of excitation at the $L_3$ and at the $L_2$, which reflects the higher degeneracy of the core state with $j = 3/2$ ($L_3$) with respect to that with $j = ½$ ($L_2$). The energy at the center of each step was chosen as the maxima of the XAS derivative around each edge, and the width of the steps was taken as the bin size (0.25 eV). This is slightly narrower than the core-hole lifetime of the Co $L_3$, which is tabulated at 0.43 eV ($\pm 25\%$) [48].

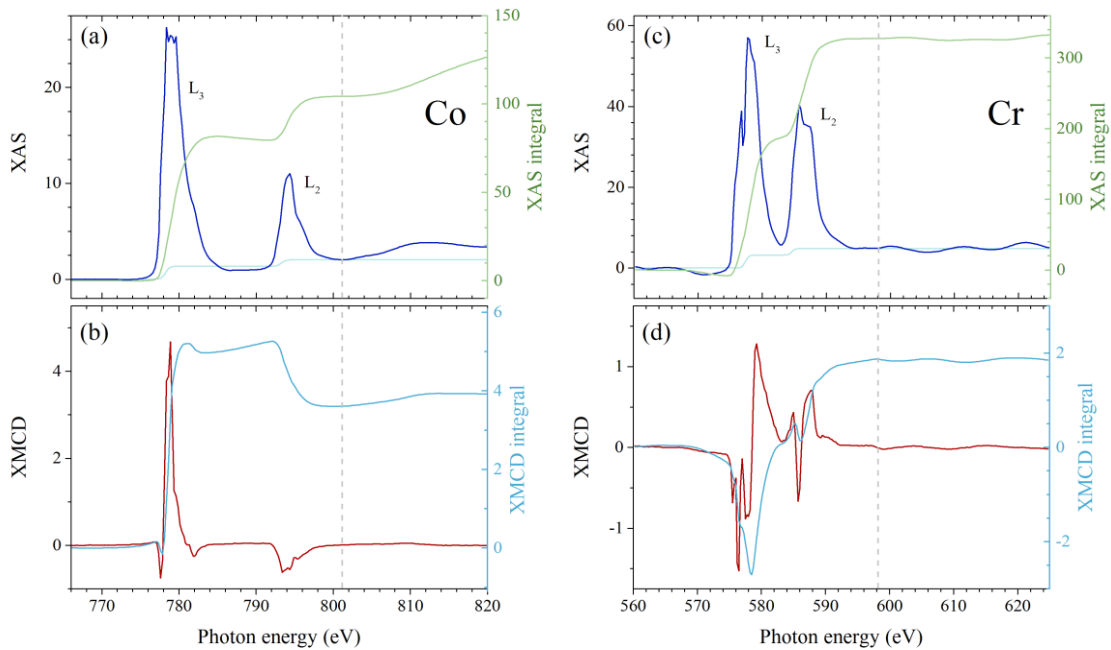

Figure 11 – XAS and XMCD spectra from CCO, taken around the Co and Cr $L_{2,3}$ edges at 15.5 K. (a) and (c) present XAS spectra ($\mu_+ + \mu_-$) around the Co and Cr L edges, respectively. The calculated two-step function is also shown (see main text). The right hand axes represent the integral over the XAS, after the two-step function is subtracted. (b) and (d) present XMCD spectra ($\mu_+ - \mu_-$) around the Co and Cr L edges, respectively. The right hand axes represent the integral over the XMCD spectra. Vertical dashed lines indicate the cutoff energies at which the integrals are evaluated (see main text).

Integrating over the XMCD spectrum allows determining the orbital moment. However, to determine the spin moment, the XMCD integral must also be evaluated over the spin-orbit split edges separately.

This is problematic in the case of Cr, because the $L_2$ and $L_3$ are so close in energy that their overlap leads to errors of up to 100% [49]. We will therefore not present spin moments for Cr. This overlap is not a severe problem for Co because the $L_2$ and $L_3$ edges are well separated, so this effect is accounted for by dividing $m_{S,eff}$ of Co by 0.92 [49]. However, choosing the correct cutoff energy between the edges can have a significant effect also on the estimate of the Co spin moment. This is demonstrated in Figure 12, which presents the calculated Co spin moment at 15.5K as a function of the cutoff energy. Also shown is the derivative of the XMCD spectrum. We choose the cutoff energy at which this derivative is zero (marked by a solid vertical line).

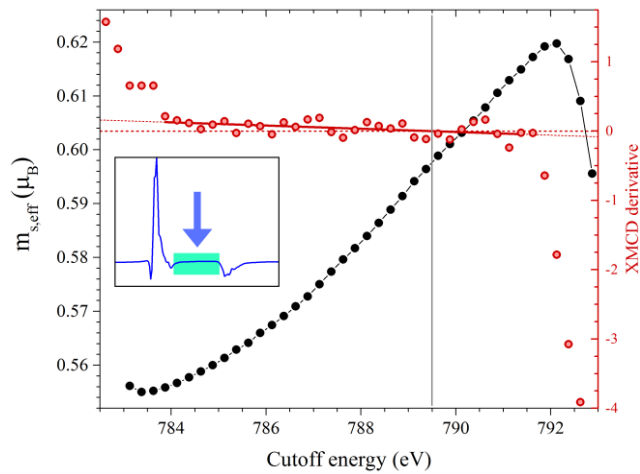

Figure 12 – Effective Co spin moment size at 15.5K, as function of the cutoff energy between the $L_2$ and $L_3$ edges. The open circles and the right hand axis indicate the derivative of the XMCD spectrum (Figure 11c). The derivative can be described by a linear function, and the diagonal line represents the best fit. The vertical line indicates the energy at which the derivative crosses zero. The inset shows the Co XMCD spectrum, highlighting the energy range shown in the main figure.

Zero-field measurements such as those in Figure 11 were taken at different temperatures, following field cooling (described above). The results of sum rules analysis of these data are presented in Figure 13, and were calculated following the analysis of Chen et al in Ref. [46]. The number of *d* holes was set in the calculations to 7 for $Cr^{3+}$ and 3 for $Co^{2+}$. Due to the field cooling procedure, we assume a single domain state along the [001] magnetic easy axis. Therefore, to account for the $\theta = 30°$ incidence angle, a factor of $\cos 30°$ was corrected for in the vertical axis. We find that the Co orbital and effective spin moments acquire the same sign. For the case of Cr these moments acquire opposite signs ($m_{S,eff}$ values are not shown for Cr). All of the moments remain roughly constant below 60 K, consistent with the temperature dependence of XMCD in scattering. The ratio $m_L/m_{S,eff}$ for Co remains around 0.3, but a slight decrease appears to occur within the multiferroic conical phase at low temperatures.

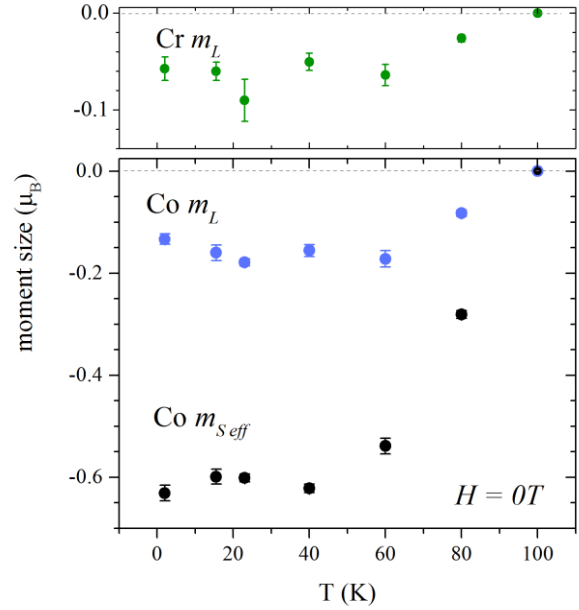

Figure 13 – Average effective spin and orbital moments ($m_{S,eff}$ and $m_L$) at various temperatures, in zero field. The vertical axis is corrected for a factor of cos 30°, to account for the angle between the beam and the [001] magnetic easy axis.

The same data were taken under a magnetic field of 6.8 T, applied parallel to the beam. The results of sum rules analysis of these data are presented in Figure 14. In this case it is unclear if the moments are still fully along the [001] or along the field direction, so the $\cos 30°$ correction was not applied in this figure (note also that the 2 K dataset was taken at 6 T, not 6.8 T). Nevertheless, the moments grow by over 100 %. As before, the moments appear mostly saturated below 60 K, and the ratio $m_L/m_{S,eff}$ for Co remains approximately 0.26.

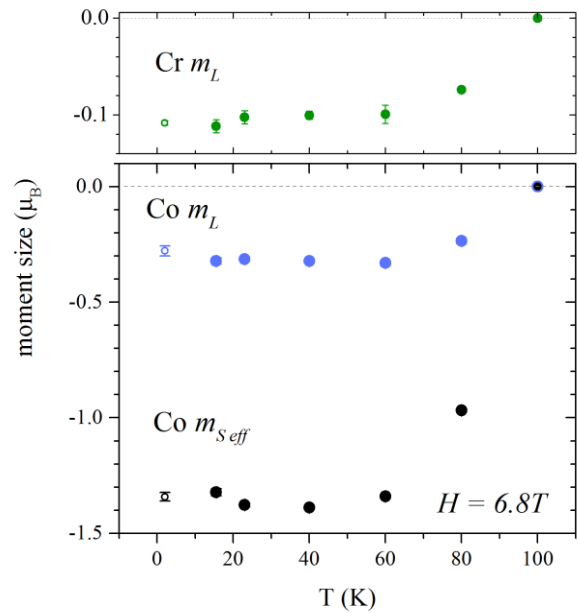

Figure 14 – Average effective spin and orbital moments ($m_{S,eff}$ and $m_L$) at various temperatures, under 6.8 T applied along the beam direction. The data points at 2K were taken under 6T, and are therefore shown as open circles.

Due to the high coercivity observed in magnetization measurements, XMCD contrast was also collected as a function of magnetic field at the Co and Cr $L_3$ edges. This allows disentangling the contribution to the hysteresis curve of magnetization from the Co and the Cr ions. For every applied field, a measurement was taken at the $L_3$ edge (576.3 and 778.9 eV) for both incoming circular polarizations. Measurements were also taken off resonance at an energy well-below the edge (574 eV and 776 eV), to correctly extract $\mu$ from Eq. (4). The Co and Cr data are presented on the same scale. As before, the beam was at a 30° angle with respect to the [001] axis, and 60° from [110]. No correction was made to account for the 30° incidence, because from these data one cannot distinguish whether the magnetization rotates or not. From the results in Panels (a)-(e) it appears that the sign of hysteresis for the Co and the Cr signals is reversed, and they appear to exhibit the same coercive field.

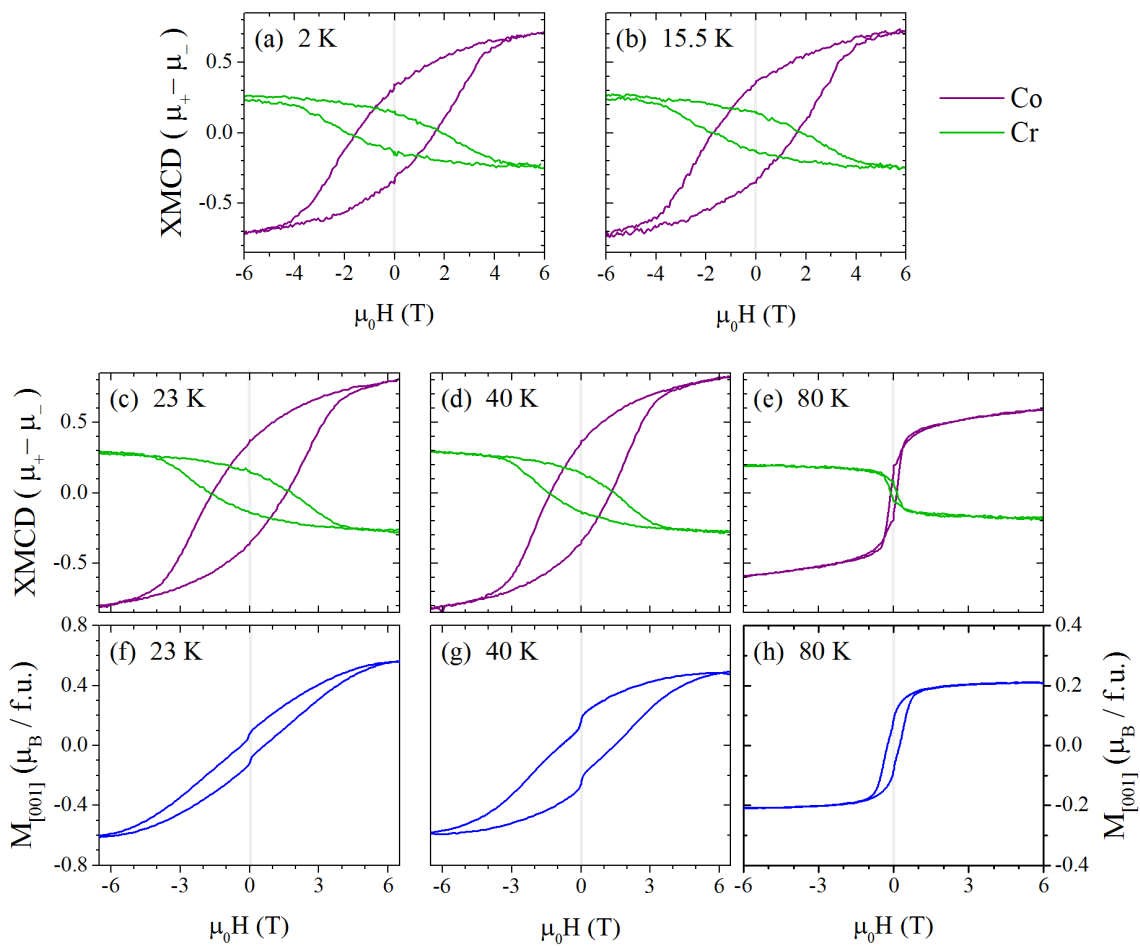

Figure 15 – Magnetic hysteresis of CCO. (a)-(e) XMCD as function of magnetic field measured with photons energies at the Co and the Cr $L_3$ edges (See main text). Data are shown for 5 different temperatures, and have been corrected using Eq. (4). All XMCD data are on the same scale. (f)-(h) Total magnetization as function of magnetic field, measured at a number of different temperatures. Panel (f) has a vertical scale than Panels (g) and (h), which have the same scale. Note that comparison between magnetization and XMCD is not straightforward because the magnetization was measured along [001], while XMCD was measured at an angle of 30° from the [001] axis.

The lower row of panels in Figure 15 presents hysteresis curves of magnetization taken with the MPMS device at selected temperatures, as previously described. A linear term has been removed from all magnetization data above 6 T (approximately the saturation field at the lowest measured temperature). This term can arise both from impurities in the substrate, as well as from non-hysteretic contributions to magnetization in the sample. The magnetization data exhibit a jump around 0 T. This cannot be reproduced by the sum of the XMCD data, because the Co and Cr signals exhibit the same coercive field. However such a comparison is only qualitative. The data from the two experiments cannot be directly compared, because the XMCD experiment was conducted at a 30° angle of incidence from the [001] axis, and magnetization measurements were taken along [001]. This also means that in the XMCD experiment the field component along [001] is weaker because it is applied along the beam direction. The discrepancy around 0 T could be explained by an unstrained fraction of the sample, which would have a low saturation field, as in bulk. This is possible because the magnetization measurement probes the whole sample, but the X-ray experiments probe only the volume beneath the 30 x 220 $\mu m^2$ beam.

## Discussion

The interest in CCO is mainly due to the macroscopically observed **M** and **P**, and the strong magnetoelectric coupling between the different magnetic sublattices. Our results provide important details about the magnetic behavior of the Co and the Cr sublattices. Even though the focus of our study is on a strained film and not bulk CCO, much of the behavior is qualitatively similar to bulk. For example the $T_S$ value is at the exact temperature as in bulk, and $T_N$ is a few Kelvin lower than bulk.

The XMCD results describe the behavior of the Co and Cr ions' average uncompensated moment. They exhibit the exact same coercive field at all temperatures, indicating that these sublattices are closely coupled to each other. Indeed the exchange coupling between them ($j_{AB}$) is understood to be either the strongest, or among the strongest interactions in CCO [50]. On the other hand, we observe opposite behaviors at the two edges: they exhibit opposite XMCD contrast as function of magnetic field (both in the sign of their hysteresis curves and in field cooling), and they have opposite signs of their spin moment (for $Co^{2+}$ the $m_L$ and $m_{seff}$ have the same sign, for $Cr^{3+}$ they are opposite). This is in agreement with the expected antiferro-type coupling between Co and Cr.

We must emphasize here that while XMCD is element-selective, it does not directly distinguish between different magnetic sublattices of the same ion species. According to LKDM theory, CCO possesses six magnetic sublattices (two of Co ions, four of Cr). This means that values obtained for either ion species represents an average over the magnetic sublattices of that species.

The shape and opposite sign of the temperature dependences of XMCD from Co and Cr can explain the temperature dependence of macroscopic magnetization in CCO. Our data are in particularly good agreement with the simple model of Ref. [13], in which the two sublattices are described by modified Brillouin functions. This model estimated a coupling constant of -18K (-1.5 meV), which is in reasonable agreement with the calculated $j_{AB}$ values of $\sim -3.5$ meV to $\sim -4.4$ meV in Refs. [50] and [41]. Given that the Co ions order at $T_C \approx 94K$, the model also predicted ordering of the Cr ions at 49K, in agreement with $T_{kink}$. However, this model requires the Co sublattice to order at $T_C$ for *any* magnetization to appear. Once magnetization of the Co sublattice exists, the model works because the Co couples to Cr. This is in contrast to the prediction that $j_{AA}$ (Co-Co coupling) is weaker than $j_{AB}$ and $j_{BB}$.

LKDM theory is a more reliable foundation to describe CCO, as it successfully reproduces many observations of AB$_2$O$_4$ magnetic Spinels. Unfortunately, this model does not correctly describe the azimuthal dependence of the ($q\ q\ 0$) reflection. For it to do so, a modification in the relation between the Co sublattices is required: a phase of $\pi/4$ between their $\gamma_\nu$ values must be introduced. This may be due to the strained state of the system. It may also underline the main shortcoming of LKDM theory, namely that it does not take into account the A-A (Co-Co) interaction. This disagreement with theory can also be the case for bulk CCO, and not due to strain. Indeed even for bulk CCO, a stability parameter value of $u = 2$ is required to reproduce the experimental results, which is outside of the expected tolerance range [9]. In addition to LKDM theory, new computational studies have been conducted on CCO, using both LSDA+U [50] and Monte Carlo simulations [51]. First-principles calculations aimed at studying the introduction of Fe ions into CCO were also conducted [41]. The magnetic interactions, as well as the magnetic ordering wave vector, were found to be directly connected to the magnitude of polarization. Most important was the finding that the effect of changing the magnitudes of the intra-sublattice couplings $j_{BB}$ and $j_{AA}$ are not negligible compared to the inter-sublattice coupling $j_{AB}$. This agrees with the observation that XMCD signals of Co and Cr appear at the same ordering temperature $T_C$, and that they exhibit the exact same coercive field at all temperatures.

The second major point in our study is the effect of strain on the system. Already from LKDM theory, the parameter $u$ was defined to quantify the level of distortion in the system. It was argued that by altering exchange paths, the Néel state destabilizes. Clearly the same idea applies for strain. Based on the notion that the $u$ parameter controls the ground state, one can intuitively realize that since $u$ depends linearly on exchange terms, strain will alter it by varying the bond lengths (and thus changing exchange paths).

Our film was grown along [110], a direction which keeps both macroscopic observables in plane. This allows applying strain directly along these directions. The first observation in relation to strain is that the multiferroic conical phase is still present (indicated by the appearance of the Co magnetic spiral), and its onset temperature remains $T_S = 27K$, as in bulk. In our case the [1$\bar{1}$0] direction of *P* remained virtually unstrained (0.02% compressive). Since the spin spiral and ***P*** are linked to the same transition, this may explain why the spiral appears unaffected by strain.

The second observation in the context of strain is that the coercive field grows by 2 orders of magnitude compared to bulk. This is already implied by the nonzero $\alpha$ term in the spin rotation energy (a consequence of a lowering of crystal symmetry from cubic), as it predicts a deepening of the energy landscape. Furthermore it is reasonable to assume that strain along a principal axis will affect the anisotropy of octahedral sites (Cr). The effect of strain on the tetrahedral sites (Co) is discussed in more detail in Ref. [17]. The sum rules analysis of the XMCD data indicates a large $Co^{2+}$ orbital moment ($m_L$). This suggests a strong magnetic anisotropy at the ionic level. The $Co^{2+}$ $m_L$ moment is not only much larger than the $Cr^{3+}$ $m_L$, is it also ~30% of the $Co^{2+}$ effective spin moment. Unfortunately no literature data is available to compare to these to bulk values, so a relation between these results and the enlarged coercivity is not directly proven here.

The possibility of an exchange bias in the system is suggested by the scattering XMCD data. This is not supported by other datasets, and is not clearly observed in the hysteresis curves of magnetization. Also, the less-understood features of magnetization in bulk do not appear in our strained sample. The $T_L$ transition is notably absent in RSXD, and no clear lock in of $q$ is observed. Strain has recently been shown to alter the lock-in transition of magnetic order in other multiferroic films [38]. Nevertheless the observed order produces a broad diffraction peak due to its short correlation length. If this phase is only at the surface of the sample, other behaviors may be buried below, but their signal is masked by the broad signal observed.

No evidence of $T_{kink}$ is observed either (both in RSXD and in XMCD), indicating that it is perhaps absent due to the strained state of our lattice. This is noteworthy because this feature is expected to relate to the Cr-Cr interaction. Such B-B interactions were understood as the main driving force behind the frustration that destabilizes the Néel state. However, the Néel state is indeed destabilized, evidenced by the appearance of the Co spiral, as in bulk. This discrepancy may be due to LKDM theory's inaccurate description of CCO. Alternatively, the Co and Cr ions may depend differently on strain directions.

Many of the present findings cannot be directly compared to bulk CCO values because of the absence of soft X-ray studies in literature. Literature on XMCD and sum rules analysis on bulk CCO would be of value, as would linear dichroism experiments. These would provide insight on the ions' local environments and on variations between differently strained states. The RSXD literature on bulk CCO does not describe the azimuthal dependence of ($q$ $q$ 0), so one cannot conclude that the disagreement with theory is a result of strain.

# Summary

Using resonant soft X-ray techniques, we observed the magnetic behavior of a strained [110]-oriented film of $CoCr_2O_4$. We used XMCD in scattering and in transmission modes to separately follow the magnetic moments of the Co and Cr ions as functions of temperature and magnetic field. Transmission measurements were facilitated by X-ray excited optical luminescence from the substrate, which allowed us to apply sum rules analysis. Resonant soft X-ray diffraction was used to follow the conical order in the multiferroic phase.

Many of the observed behaviors are close to bulk behavior, even though the system is strained. For example, the sign of the effective spin moments ($m_{S,eff}$) of $Co^{2+}$ and $Cr^{3+}$ is opposite, as expected to occur also in bulk. The XMCD signal of the two sublattices is always of opposite sign. The temperature dependence of XMCD from the two sublattices is qualitatively different, also as predicted for bulk.

Two major effects of strain are observed: the first is is that the coercive magnetic field is 2 orders of magnitude higher than in bulk. This is well-described by a quadratic term ($\alpha$) in the spin rotation energy, which appears due to symmetry lowering. Sum rules analysis of the XMCD data indicates a large $Co^{2+}$ orbital moment ($m_L$), which suggests a strong single-ion magnetic anisotropy. The second observation is that the conical magnetic structure appears at the same $T_S$ temperature as in bulk, demonstrating that the low temperature phase is not suppressed by the strain. This order is strongly coupled to the electric polarization, so its appearance is a strong indication that this phase is multiferroic despite the applied strain. The conical order differs from bulk because the lock-in transition at $T_L \approx 14.5\ K$ is not observed, and because the azimuthal dependence of the $(q\ q\ 0)$ reflection cannot be described by LKDM theory. The azimuthal dependence of $(q\ q\ 0)$ can be described by slightly modifying the Co spin motif in the LKDM model.

The present work serves to underline the prospect of manipulating ferrimagnetic multiferroics using strain. We have demonstrated that the high degree of magnetic frustration can be manipulated to alter the functionality of *M*. It remains to be seen whether or not ***P*** can be manipulated in a similar fashion, because measurements of ***P*** in CCO films have not been reported to date.

# Acknowledgements

We gratefully thank the X11MA, X07MA and X04SA beam line staff for experimental support. The financial support of PSI, the Swiss National Science Foundation. E.M.B. acknowledges funding from the European Community's Seventh Framework Programme (FP7/2007-2013) under grant agreement n.°290605 (PSI-FELLOW/COFUND).